\documentclass[prb,twocolumn,amsmath,amssymb]{revtex4-2}

\usepackage{graphicx}
\usepackage{dcolumn}
\usepackage{bm}
\usepackage{mathrsfs}
\usepackage{subfigure}
\usepackage{natbib}
\usepackage{amsmath}
\usepackage{amssymb}
\usepackage{Jonasmacros}
\usepackage{mathptmx}
\usepackage{txfonts}
\usepackage{ifsym}
\usepackage[usenames,dvipsnames,svgnames]{xcolor}

\usepackage[colorlinks,citecolor=magenta,linkcolor=red]{hyperref}
\usepackage[usenames,dvipsnames,svgnames]{xcolor}

\begin{document}
\date{\today} 

\title{Vibrationally Induced Magnetism in Supramolecular Aggregates}

\author{J. Fransson}
\email{Jonas.Fransson@physics.uu.se}
\affiliation{Department of Physics and Astronomy, Box 516, 75120, Uppsala University, Uppsala, Sweden}



\begin{abstract}
Magnetic phenomena are in chemistry and condensed matter physics considered to be associated with low temperatures. That a magnetic state, or order, is stable below a critical temperature as well as becoming stronger the lower the temperature is a nearly unquestioned paradigm. It is, therefore, surprising that recent experimental observation made on supramolecular aggregates suggest that, for instance, the magnetic coercivity may increase with increasing temperature, as well as the chiral induced spin selectivity effect may be enhanced. Here, a mechanism for vibrationally stabilized magnetism is proposed and a theoretical model is introduced with which the qualitative aspects of the recent experimental findings can be explained. It is argued that anharmonic vibrations, which become increasingly occupied with increasing temperature, enables nuclear vibrations to both stabilize and sustain magnetic states. The theoretical proposal, hence, pertains to structures without inversion and/or reflection symmetries, for instance, chiral molecules and crystals.
\end{abstract}

\maketitle

Magnetism and magnetic phenomena concern central questions in physics and chemistry and remain to be major fields of research. The phenomenology of magnetism pertains to seemingly separated questions of, for instance, room temperature ferromagnetism \cite{NatPhysics.5.840,NatNanotech.13.289,JACS.140.11519,APL.92.082508,APL.81.4212,NanoLett.9.220}, topological matter \cite{Science.329.61,Nature.546.270,NatMaterials.19.484,Nature.603.41}, and exotic spin excitations (e.g., skyrmions) \cite{Nature.442.797,Science.323.915,Nature.465.901,ChemRev.121.2857}. Furthermore, results from basic science of magnetism have led to important applications in, for example, medicine (e.g., magentic resonance imaging, MRI) \cite{Nature.242.190,PhysRevB.12.3618} and data storage (hard drives based on giant magneto resistance) \cite{PhysRevLett.61.2472,PhysRevB.39.4828}. Recent studies also suggest that magnetic phenomena are not only pertinent in biological contexts but are also crucial for, e.g., oxygen redox reactions on which aerobic life depends \cite{PNAS.2202650119,PNAS.2204765119,Gupta2022.11.29.518334}.

Ordered magnetic states typically form below a critical temperature, $T_c$. Above this temperature, higher energy excitations become occupied, due to thermal fluctuations, which tend to have a randomizing, hence, detrimental effect on the magnetic order. An overwhelming amount of evidence for this behavior has been collected over the years, which naturally has also led to the idea that magnetic order is perceived as a low temperature collective phase, albeit $T_c$ may be larger than 1,000 K in some compounds \cite{Keffer1966,JPhysD.49.095001}.
Therefore, from this point of view, the recent reports on supramolecular aggregates that show stable room temperature ferromagnetism, however, small, if not vanishing, magnetic moment at low \cite{JPhysChemLett.7.4988,ACSNano.14.16624}, are indeed surprising. Nevertheless, while the presented measurements suggest an increasing coercive field with temperature, one would, at least from an entropy argument, still have to expect the existence of a critical temperature above which the magnetic order ceases to exist.

As a destructive mechanism for magnetic order, nuclear vibrations and their collective correspondence, phonons, are considered as a source for decoherence and dissipation of stable magnetic states. It is, therefore, easy to make a connection between increased temperatures and phonon activation which leads to energy level broadening, occupation of multiple electronic states with competing magnetic properties, and distortions and reformations of nuclear configurations. Hitherto, nuclear vibrations have never been considered as a source for stabilizing magnetic order.

In this article, a coupling between spin and vibrational degrees of freedom that provides a mechanism for stabilizing finite magnetic moments in molecular aggregates and crystals is proposed. The mechanism is shown to generate an increasing magnetic moment with increasing temperature as well as a critical temperature below which the magnetic moment becomes small, or, negligible. The source for this property lies in the nature of the vibrational excitations. For sufficiently low temperatures, the nuclear vibrations are dominated by harmonic oscillations around some equilibrium position which neither charge nor spin polarize the structure. For higher temperatures, anharmonic vibrational modes are occupied which may cause a net average nuclear displacement from the equilibrium position. Hence, this displacement leads to a local charge polarization which exert a force on the local spins. Then, should the compound lack inversion and/or reflection symmetry, a net magnetic moment can be developed and maintained.

Simply described, the effect can be viewed in an dimer of vibrating spins, with spin $S=1/2$, which are coupled via indirect exchange through an electronic medium \cite{PhysRevMaterials.1.074404}. Then, while an isotropic spin exchange generates the set of singlet and triplet states, a symmetric anisotropic spin exchange separates the triplet into a singlet and a doublet. Finally, via a spin-vibration coupling the nuclear vibrations provide a pseudo-magnetic field, which spin-splits the remaining doublet. Regardless of whether the isotropic spin exchange favors a ferromagnetic or anti-ferromagnetic zero temperature ground state, the excited states begin to become occupied whenever thermal excitations are available as the temperature rises. Then, the occupation of the spin-split doublet is unequally distributed between the two states, which results in a finite magnetic moment of the dimer.

The question to be addressed here is whether nuclear (ionic, molecular) vibrations can stabilize magnetic configurations or a magnetic order in structures in which the time-inversion symmetry is sustained in the absence of such vibrations. The investigation is stimulated by the experimental observations of room temperature ferromagnetism in self-assembled ensembles of paramagnetic and diamagnetic supramolecular aggregates or other types of molecular structures \cite{JPhysChemLett.7.4988,JPhysD.49.095001,ACSNano.14.16624,JPhysChemC.121.12159,Science.370.587,JPhysChemC.125.9875,ACSApplMaterInterfaces.13.34962,JACS.144.7302}.


The discussion is begun by considering the introduced example of the spin dimer $\bfS_1\otimes\bfS_2$ which, albeit simplified, makes it possible to highlight some of the generic aspects of the proposed mechanism. Therefore, the underlying physics can be illustrated by the model Hamiltonian \cite{PhysRevMaterials.1.074404}
\begin{align}
\Hamil_\text{MQ}=&
	-
	J\bfS_1\cdot\bfS_2
	-
	\sum_{mn}
		\Bigl(
			S_m^zI_{mn}S_n^z
			+
			S_m^z\calA_{mn}\cdot\bfQ_n
		\Bigr)
	,
\label{eq-TwoHalfsModel}
\end{align}
where $\bfS_m$ and $\bfQ_m$, $m=1,2$, denote the spin moment and nuclear displacement operator vectors, respectively. The spins are mutually coupled via the isotropic, $J$, and symmetric anisotropic, $I_{mn}$, spin exchanges, as well as coupled to the nuclear displacements through the spin-lattice coupling $\calA_{mn}$. Here, for the sake of clarity, however, without loss of generality, only the $z$-projections of the spins have been included in the anisotropic and spin-lattice couplings.

The spectrum of this model is given by the four energies
\begin{align}
E^{(-)}_\pm=
		-\frac{J}{4}-I_+\pm\alpha
		&,&
E^{(+)}_\pm=&
	\frac{J}{4}-I_-\pm\Omega
	,
\end{align}
where the parameters $I_+=\sum_{mn}I_{mn}/4$, $I_-=I_+-(I_{12}-I_{21})/2$, $\alpha=\sum_{mn}\calA_{mn}\cdot\av{\bfQ_n}/2$, and $\Omega=\sqrt{J^2/4+\zeta^2}$, in which $\zeta=\alpha-\sum_m\calA_{2m}\cdot\av{\bfQ_m}$, whereas $\av{X}$ denotes the expectation value of an operator $X$.

\begin{figure}[t]
\begin{center}
\includegraphics[width=\columnwidth]{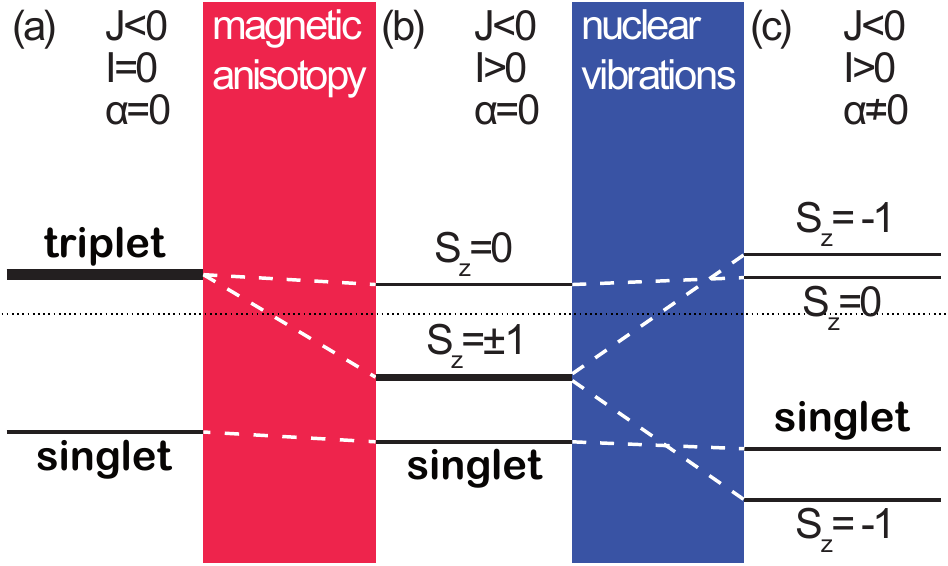}
\end{center}
\caption{Illustration of a possible energy diagram for (a) the anti-ferromagnetically coupled spin dimer under influence of (b) easy axis anisotropy and (c) nuclear vibration.}
\label{fig-TwoHalfsEnergies}
\end{figure}

The possible outcomes of this spectrum, is conveniently analyzed by first considering all parameters but $J$ to vanish. Then, the solutions can be categorized into the singlet ($E_S=3J/4$, $\ket{S=0,m_z=0}$) and triplet ($E_T=-J/4$, $\ket{S=1,m_z=m}$, $m=0,\pm1$) states, with a singlet (triplet) ground state whenever the isotropic spin exchange is anti-ferromagnetic, $J<0$, (ferromagnetic, $J>0$). An illustration of this scenario is shown in Fig. \ref{fig-TwoHalfsEnergies} (a), for an anti-ferromagnetic spin exchange.

Starting from the anti-ferromagnetic ground state, $J<0$, the introduction of vibrations may result in a strong change of the spin ground state. By adding the magnetic and vibrational contributions to the spin excitations, the degeneracies of the states break up and the corresponding energies shift according to the scheme laid out in Fig. \ref{fig-TwoHalfsEnergies}. Structurally, the negative spin exchange $J$, leads to that the singlet state is the ground state, in absence of $I_\pm$, $\alpha$, and $\zeta$, whereas the triplet states represent the excited states. While addition of the magnetic anisotropies, in principle, may lower the energies of the states $\ket{S=1,m_z=\pm1}$ below the singlet, it is not likely to happen since the amplitude of the parameters $I_\pm$ are often, but not always, significantly smaller than $|J|$  \cite{PhysRevMaterials.1.074404}. By contrast, the energy scales set by $\alpha$ and $\zeta$, which are attributed to the spin-lattice coupling, may assume values with amplitudes well within the order of $|J|$, and also larger \cite{PhysRevMaterials.1.074404}. Then, this may lead to a changed ground state from the low spin ($\av{\bfS_1\otimes\bfS_2}=0$) to a high spin ($\av{\bfS_1\otimes\bfS_2}\neq0$) configuration. The requirements for such a transformation can be formulated in terms of the condition $\zeta^2<-(|J|-\calI-|\alpha|)(\calI+|\alpha|)$, where $\calI=I_+-I_-$. This condition opens up the possibility that it would be sufficient with a small $\zeta$, which occurs whenever $\calI+|\alpha|\gtrsim|J|$. Physically, this means that the spin-lattice couplings $\calA_{1m}$ and $\calA_{2m}$ are not required to be asymmetric for the nuclear vibrations to sustain the magnetic ground state.

In a more generalized form, one may assume that the localized spin moments, as well as the nuclear motion, are embedded in an electronic structure which constitutes the spin-spin and spin-vibration couplings \cite{PhysRevMaterials.1.074404}. In such environment, it can be shown that the spin-lattice coupling is non-vanishing whenever the background electronic structure has either a viable spin density and/or a spin-orbit coupling. Hence, while the former requirement implies a preexisting ordered magnetic state of the structure, the latter does not. In the following it will merely be assumed that the spin-lattice coupling is non-zero, while the specific origin for this state is non-essential. Without assuming a specific form of a general lattice in which the spin and vibrational components are embedded, the considered structure can be summarized in a model of the form
\begin{align}
\Hamil=&
	\Hamil_S
	+
	\Hamil_\text{vib}
	-
	\sum_{mn}J_{mn}\bfS_m\cdot\bfS_n
	+
	\sum_{mn}\bfS_m\cdot\bfA_{mn}Q_n
	.
\label{eq-model}
\end{align}
Here, the first two terms include the local on-site spin (anisotropy) and vibrational (modes, anharmonicity) properties, respectively. The third term describes the spin-spin interactions between the spin moments, whereas the last term accounts for spin-vibration interactions in the lattice. The displacement operator $Q_m$ is expressed in terms of the vibrational creation and destruction operators $a_m$ and $a^\dagger_m$, respectively, such that $Q_m=a_m+a^\dagger_m$.

\begin{figure*}[t]
\begin{center}
\includegraphics[width=\textwidth]{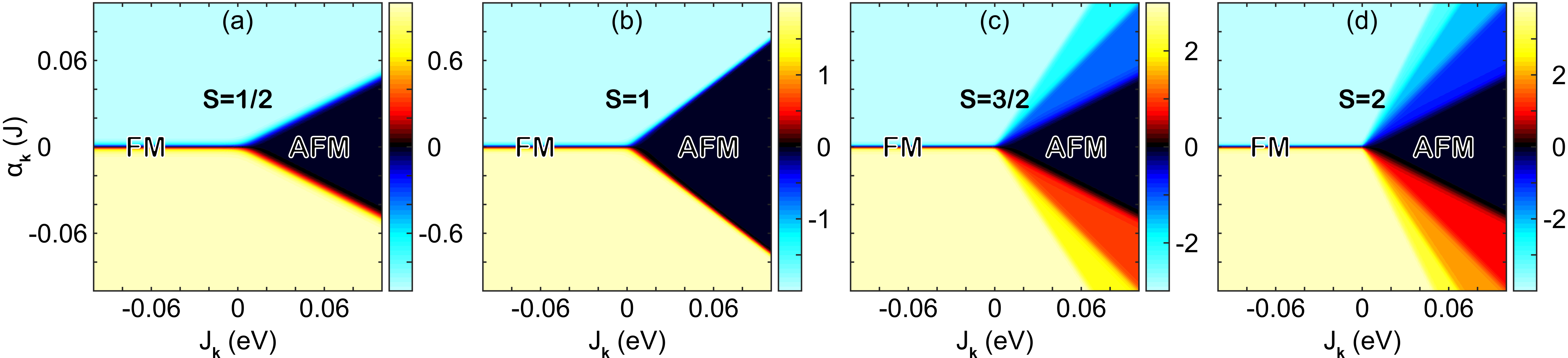}
\end{center}
\caption{Phase diagrams for spin dimers $\bfS_\bfk\otimes\bfS_{\bar\bfk}$ with spin (a) $S=1/2$, (b) $S=1$, (c) $S=3/2$, and (d) $S=2$ as function of the spin exchange $J_\bfk$ and nuclear displacement field $\alpha_\bfk=A_\bfk\av{Q_{\bar\bfk}}$. The color scale denotes the total spin moment of the dimer, where black signifies the anti-ferromagnetic regime, whereas the colored areas correspond to magnetized ferromagnetic regimes. Here, the temperature $T=3$ K. The ranges for ferromagnetic (FM, $J_\bfk<0$) and anti-ferromagnetic (AFM, $J_\bfk>0$) spin exchange are indicated.}
\label{fig-PhaseDiagram}
\end{figure*}

The following investigation is focused on the magnetic moment $\av{\bfS_m}$ per site, as well as the collective magnetization $\av{\bfM}=\sum_m\av{\bfS_m}$ in the whole structure.
%
In reciprocal space, the spin operator $\bfS_m=\sum_\bfk\bfS_\bfk e^{-i\bfk\cdot\bfr_m}/\sqrt{N}$, where $N$ is the number of sites lattice, and analogously for the vibration operators $a_m$. The diagonal structure of $\Hamil_S$ is preserved also in reciprocal space under the assumption that $E_{m\alpha}=E_\alpha$ for all $m$, which is justified under the assumption that, in the absence of interactions, the local environment of each spin moment is equivalent. It is equally justified to assume that the local vibrational modes are the same for all molecules of the same kind. Therefore, the Hamiltonian for the a single vibrational mode is given by $\Hamil_\text{vib}=\sum_\bfq\omega_0a^\dagger_\bfq a_\bfq+\Hamil_\text{anharm}$, where the last term contains possible anharmonic properties of the vibrational structure.

The Heisenberg model, $-\sum_{mn}J_{mn}\bfS_m\cdot\bfS_n$, is simplified by assuming that the exchange parameter is distance dependent on the simple form $J_{mn}=J(\bfr_m-\bfr_n)$, such that $J_\bfk=-\sum_mJ(\bfr_m-\bfr_n)e^{i\bfk\cdot(\bfr_m-\bfr_n)}=-\sum_mJ(\bfr_m)e^{i\bfk\cdot\bfr_m}$, which is independent of the neighboring sites $n$. Then, the spin-spin interactions can be written as $-\sum_{mn}J_{mn}\bfS_m\cdot\bfS_n=\sum_\bfk J_\bfk\bfS_\bfk\cdot\bfS_{\bar\bfk}$, where $\bar\bfk=-\bfk$. By the same token, the interaction term is written $\sum_m\bfS_m\cdot\bfA_{mn}Q_n=\sum_\bfk\bfS_\bfk\cdot\bfA_\bfk Q_{\bar\bfk}$, with $Q_\bfk=a_\bfk+a^\dagger_{\bar\bfk}$, and $\bfA_\bfk=\sum_n\bfA_n e^{i\bfk\cdot(\bfr_m-\bfr_n)}$. It should be noticed that $\bfA_0=0$ since there is no net displacement of the compound.
The model is in reciprocal space, hence, reformulated to read
\begin{align}
\Hamil=&
	\Hamil_S+\Hamil_\text{vib}
	+
	\sum_\bfk\bfS_\bfk\cdot
		\Bigl(
			J_\bfk\bfS_{\bar\bfk}
			+
			\bfA_\bfk Q_{\bar\bfk}
		\Bigr)
		.
\label{eq-Hamil_k}	
\end{align}
It should be noticed that this formulation is general and applies equally to orded and disordered compounds, since the derivation does not rely on any assumption about a regular lattice structure.

\begin{figure}[b]
\begin{center}
\includegraphics[width=0.5\columnwidth]{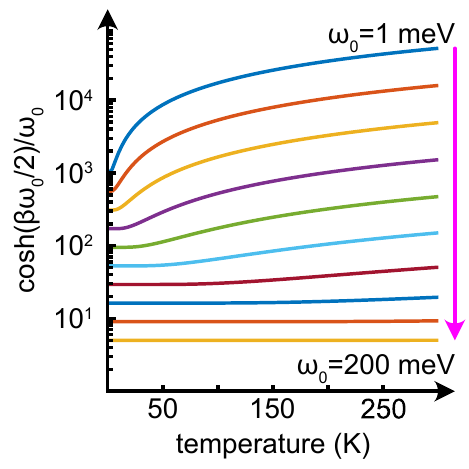}
\end{center}
\caption{Temperature dependence of $\coth(\beta\omega_0/2)/\omega_0$ for $\omega_0\in\{$1.0, 1.8, 3.2, 5.8, 10.5, 19.0, 34.1, 61.5, 111, 200\} meV.}
\label{fig-AT}
\end{figure}

The conversion of the model into reciprocal space, taken into account the simplifications, leads to that the argument applied for the spin dimer above can be utilized for each momentum $\bfk$. Indeed, for each $\bfk$, the mean field description, $Q_\bfk\rightarrow\av{Q_\bfk}$, leads to that the pair of spins $\bfS_\bfk$ and $\bfS_{\bar\bfk}$, depending on $J_\bfk$, forms either a singlet or triplet ground state in the limit $\bfalpha_\bfk=\bfA_\bfk\av{Q_{\bar\bfk}}\rightarrow0$, whereas the degeneracy of the triplet state breaks for non-vanishing $\bfA_\bfk\av{Q_{\bar\bfk}}$, c.f., Fig. \ref{fig-TwoHalfsEnergies}.

For a spin-lattice coupling $\bfA_k=A_\bfk\hat{\bf z}$, the spectrum of the spin-dimer $\bfS_\bfk\otimes\bfS_{\bar\bfk}$ corresponding to $\Hamil_\bfk=\bfS_\bfk\cdot(J_\bfk\bfS_{\bar\bfk}+\alpha_\bfk\hat{\bf z})$, is for spin $1/2$ given by $E_\pm^{(-)}(\bfk)=(J_\bfk\pm3\alpha_\bfk)/4$ and $E_\pm^{(+)}(\bfk)=-J_\bfk/4\pm\Omega_\bfk/2$, where $\Omega_\bfk=\sqrt{J_\bfk^2+\alpha_\bfk^2/4}$. The spin moment for the spin dimer, here calculated using $\av{\bfS_\bfk\otimes\bfS_{\bar\bfk}}=\tr e^{-\beta\Hamil_\bfk}(\sigma^0\otimes\sigma^z+\sigma^z\otimes\sigma^0)\hat{\bf z}/2\tr e^{-\beta\Hamil_\bfk}$, is, therefore, given by
\begin{align}
\av{\bfS_\bfk\otimes\bfS_{\bar\bfk}}=&
	-\frac{e^{-\beta J_\bfk/4}\sinh(3\beta\alpha_\bfk/4)}
		{e^{-\beta J_\bfk/4}\cosh(3\beta\alpha_\bfk/4)+e^{\beta J_\bfk/4}\cosh(\beta\Omega_\bfk/2)}
	\hat{\bf z}
	,
\end{align}
where $1/\beta=k_BT$ is the thermal energy in terms of the temperature $T$ and the Boltzmann constant $k_B$. The spin moment is vanishingly small for positive $J_\bfk$ whenever $3|\alpha_\bfk|<J_\bfk$, that is, the anti-ferromagnetic regime prevails for sufficiently small spin-lattice coupling. Reversely, the spin dimer acquires a net moment for $3|\alpha_\bfk|>J_\bfk$ for both ferromagnetic and anti-ferromagnetic spin exchange $J_\bfk$. Moreover, this net moment changes sign with $\alpha_\bfk$, as would be expected for a field which is formally similar to a magnetic field.

The $\hat{\bf z}$-projected spin moment for the spin dimer $\bfS_\bfk\otimes\bfS_{\bar\bfk}$, $\bfk\neq0$, is plotted in Fig. \ref{fig-PhaseDiagram} for set-ups with spin (a) $S=1/2$, (b) $S=1$, (c) $S=3/2$, and (d) $S=2$, and vanishing on-site anisotropy. The plots very well illustrate the vibrationally stabilized magnetic moments of the structures for non-vanishing $\alpha_\bfk$, whenever the spin exchange $J_\bfk<0$. The transition from the zero moment anti-ferromagnetic state, $J_\bfk>0$, to a stabilized moment for sufficiently strong at the crossover condition $3|\alpha_\bfk|=J_\bfk$ is conspicuous. These calculations are consistent with the implications that was drawn for the dimer, c.f., Fig. \ref{fig-TwoHalfsEnergies}.

\begin{figure*}[t]
\begin{center}
\includegraphics[width=\textwidth]{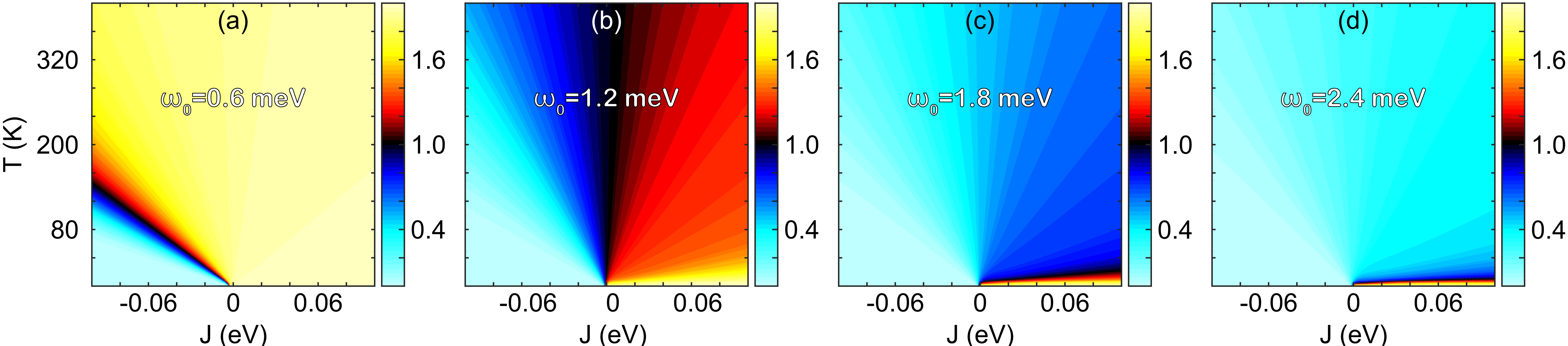}
\end{center}
\caption{Phase diagrams for spin dimers with spin $S=1$ and vibrational modes (a) $\omega_0=0.6$ meV, (b) $\omega_0=1.2$ meV, (c) $\omega_0=1.8$ meV, and (d) $\omega_0=2.4$ meV, as function of the spin exchange $J$ and the temperature $T$. The color scale denotes the total spin moment of the dimer. Other parameters are $A^{(z)}_\bfk\Phi=0.2$ $\mu$eV}
\label{fig-PhaseDiagramJvsT}
\end{figure*}

The phenomenological basis that the magnetic moment may be stabilized by the nuclear vibrations is thereby set. The next task is, then, to consider a mechanism which can support this phenomenological description.

The above discussion is based on that the mean displacement $\av{Q_\bfk}$ is non-vanishing. For simplicity and in order to study the effect of the mean displacement, consider the linear response expansion; $\av{A(t)}=\av{A}_0-i\int_{-\infty}^t\av{\com{A(t)}{\Hamil'(t')}}_0dt'+\cdots$, with respect to a perturbation $\Hamil'(t')$. To the first non-trivial order in terms of anharmonic processes $\Hamil_\text{anharm}$, the mean displacement is, then, given by
\begin{align}
\av{Q_\bfk(t)}\approx&
	(-i)
	\int_{-\infty}^t
	\Biggl(
		\sum_\bfp
			\av{\com{Q_\bfk(t)}{\bfS_\bfp\cdot(J_\bfp\bfS_{\bar\bfp}+\bfA_\bfp Q_{\bar\bfp})(t')}}
\nonumber\\&
		+\av{\com{Q_\bfk}{\Hamil_\text{anharm}(t')}}
	\Biggr)
	dt'
	.
\label{eq-linearQ}
\end{align}
Then, for a vibrational back ground that constitutes only harmonic oscillations, that is, $\Hamil_\text{anharm}=0$, the last contribution in the above expansion vanishes. Under such conditions, the mean displacement and local spin moments are interdependent, such that $\av{Q_\bfk}\neq0$ if and only if $\av{\bfS_\bfk}\neq0$. Hence, harmonic nuclear vibrations may not by themselves lead to a breaking of the a time-reversal symmetry. Harmonic nuclear vibrations may, nevertheless, facilitate a strengthening of an already existing magnetic state through a self-consistent build up of the nuclear displacement field.

By contrast, the presence of anharmonic vibrations in the structure, time-reversal symmetry may be broken. For instance, using the model $\Hamil_\text{anharm}=\sum_{\bfk\bfk'}\Phi_{\bfk\bfk'}Q_\bfk Q_{\bfk'}Q_{\bar\bfk+\bar\bfk'}$, which corresponds to the lowest order anharmonic correction to phonon Hamiltonian, one obtains from the last term in Eq. \eqref{eq-linearQ} the correction
\begin{align}
	-
	\frac{6\Phi_{\bfk\bfk/2}}{\omega_0}
	\coth\frac{\beta\omega_0}{2}
	.
\end{align}
The temperature dependence of this function is displayed in Fig. \ref{fig-AT} for vibrational energies $\omega_0$ ranging between 1 meV and 200 meV, assuming that $\Phi_{\bfk\bfk'}$ varies slowly with the temperature. It may be literally seen that the temperature variations is very strong for low energies but wanes with increasing energy. This property suggests that slow vibrations, e.g., coherent vibrations which incorporate a larger portion of the structure, can be expected to be crucial for the vibrationally assisted formation of the magnetic order. Reversely, high frequency vibrations of the nuclei may not contribute to the formation of the magnetic order but may instead be expected to be destructive for the same.

The anharmonic correction does lead to the expected properties for the spin pair $\av{\bfS_\bfk\otimes\bfS_{\bar\bfk}}$, which can be seen in Fig. \ref{fig-PhaseDiagramJvsT}. Here, the phase diagrams for the total spin of such pair with spin $S=1$ are plotted as function of the spin exchange $J$ and temparature $T$, for the vibrational energies (a) $\omega_0=0.6$ meV, (b) $\omega_0=1.2$ meV, (c) $\omega_0=1.8$ meV, and (d) $\omega_0=2.4$ meV. The four examples provide a good range of the possible magnetic properties that could be expected. For small vibrational energies, for instance, the spin--phonon coupling maintains the ferromagnetic state in a wide temperature range. It, moreover, induces the transition between the anti-ferromagnetic to a ferromagnetic state for large enough temperature. Then, for increasing vibrational energies, the ferromagnetic state is partially developed throughout the phase diagram but is fully developed at low temperature in the ferromagnetic regime only. For sufficiently large vibrational energy, finally, the ferromagnetic state is confined to low temperature in the ferromagnetic regime only.

\begin{figure}[t]
\begin{center}
\includegraphics[width=\columnwidth]{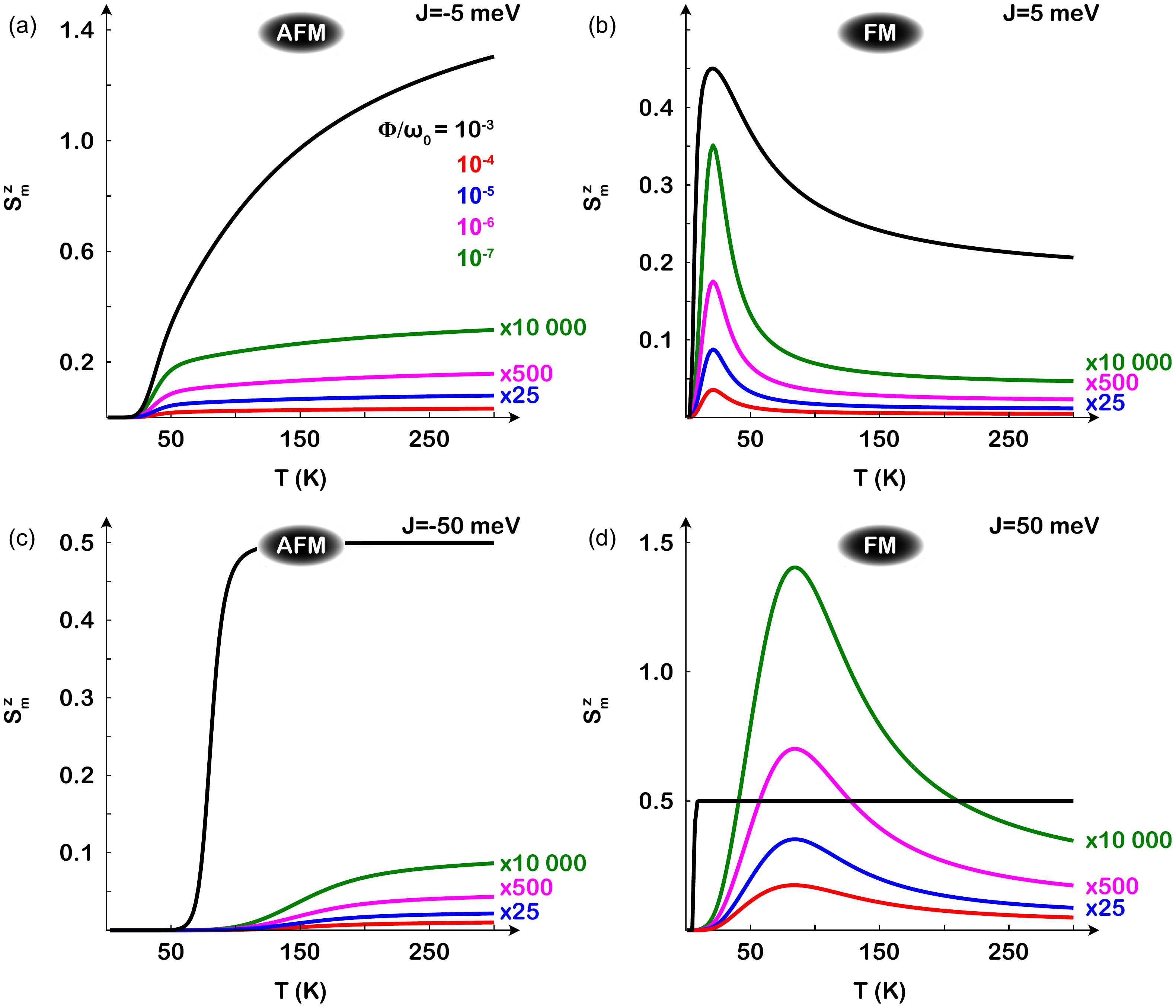}
\end{center}
\caption{Magnetic moment per site in a configuration with (a), (c) anti-ferromagnetic and (b), (d) ferromagnetic spin exchange $J$ with spin $S=1$, as function of the temperature $T$ for a few different strengths of the anharmonic coupling parameter $\Phi\in\{10^{-7},\ 10^{-6},\ 10^{-5},\ 10^{-4}, 10^{-3}\}\omega_0$, with $\omega_0=100$ $\mu$eV. The spin exchange is (a), (b) $|J|=5$ meV, and (c), (d) $|J|=50$ meV. Here, $I=|J|$, $r_S=r_A=1$ \AA. The green, magenta, and blue lines are multiplied by 10,000, 500, and 25, respectively, for visibility within the scale.}
\label{fig-MTemp}
\end{figure}

Finally, in order to obtain an estimate of the magnetic moment $\av{\bfS_m}$, hence, the magnetization $\av{\bfM}$, assume, without loss of generality, a two-dimensional structure in which the spin exchange interaction $J(\bfr_m-\bfr_n)$ can be approximated by a Gaussian distribution function, that is, $J(\bfr_m-\bfr_m)\approx Je^{-(\bfr_m-\bfr_n)^2/r_S^2}$, where $J$ is a constant energy whereas $r_S$ is the spatial decay rate. With these assumptions, the $\bfk$-dependent exchange parameter $J_\bfk$ is approximately given by $J_\bfk=-\sqrt{2}\pi r_SJe^{-k^2/k_S^2}$, where $k_S=2/r_S$. Similarly, the spin-lattice coupling $\bfA_\bfk$ can be approximated by $\bfA_\bfk=\sqrt{2}\pi r_A\bfA e^{-k^2/k_A^2}$, where $k_A=2/r_A$ in terms of the spatial decay rate $r_A$. Assuming that the coupling parameter $\Phi_{\bfk\bfk'}\approx\Phi$, that is, varies slowly with the momentum, the spin moment per site is calculated numerically. The plots in Fig. \ref{fig-MTemp} illustrate the magnetic moment per site, $\av{\bfS_m}$, in a configuration with (a), (c) anti-ferromagnetic ($J<0$) and (b), (d) ferromagnetic ($J>0$) spin exchange and spin $S=1/2$. Here, an easy plane anisotropy $I=|J|$ has been introduced in order to stabilize the anti-ferromagnetic state at low temperatures for $J<0$ as well as the corresponding superparamagnetic state for $J>0$.

A conclusion that can be drawn from the plots in Fig. \ref{fig-MTemp} is that a spin moment stabilize under the influence from nuclear vibrations. It can, moreover, be observed that increasing temperature is not necessarily detrimental for the formation and maintenance of the spin moment. Indeed, the configuration with anti-ferromagnetic spin exchange, the moment grows stronger with temperature throughout the whole range that is considered. The configuration with ferromagnetic spin exchange initially peaks and then decays towards, what appears to be, a non-zero high temperature limit. It is also important to notice that is suffices for the anharmonic coupling parameters $\Phi$ to be fairly weak, in the order of $\mu$eV, and yet make a significant difference for the magnetic properties. This observation, therefore, suggests that the vibrationally induced magnetism may be present quite abundantly in structures in which inversion and/or reflection symmetries are lacking.

It can be concluded from this discussion, that anharmonic vibrations in the structure may lead to a non-vanishing mean displacement which, in turn, can be the source of stabilization of magnetic moments. However, in a structure with inversion and/or reflection symmetries, the effects of the anharmonicity at different positions would be expected to cancel one another, partially or completely, which means that strong observable effects should be sought in compounds which lack such symmetries. In chiral structures, for example, anharmonic contributions would be expected to not cancel out due to the intrinsic absence of inversion and reflection symmetries. For instance, anharmonic contributions were proposed to increase the chiral induced spin selectivity effect with temperature in azurin \cite{JPhysChemC.125.9875}.

The results obtained within this article should be set in perspective with the recent observations of, e.g., increasing coercivity or enhanced chiral induced spin selectivity effect with temperature \cite{JPhysChemLett.7.4988,JPhysD.49.095001,ACSNano.14.16624,JPhysChemC.125.9875,ACSApplMaterInterfaces.13.34962,JPhysChemC.126.3257,JACS.144.7302}.
For instance, the molecular crystals investigated in Refs. \citenum{JPhysChemLett.7.4988,JPhysD.49.095001,ACSNano.14.16624,ACSApplMaterInterfaces.13.34962} show an increasing coercive field up to 300 K, however, accompanied by decreasing remanent and saturated magnetizations. Such behavior is reproduced by the model for ferromagnetic spin exchange, see Fig. \ref{fig-MTemp} (b). First principles calculations of the spin exchange indicate that a ferromagnetic order is favored in the compounds investigated in  Refs. \citenum{JPhysChemLett.7.4988,ACSNano.14.16624} at $T=0$ K. In the latter article, however, an anti-ferromagnetic state was reported to be nearly degenerate with the ferromagnetic state. Nuclear vibrations have, moreover, previously been shown the make a significant difference for the theoretical modeling of the chiral induced spin selectivity effect, see, e.g., Refs. \citenum{JPhysChemC.125.9875,JPhysChemC.126.3257,PhysRevB.102.235416,NanoLett.21.3026,JPhysChemLett.13.808,JACS.144.7302}.

It has been shown that by the existence of a coupling between the electronic spin and nuclear vibrations, nuclear vibration may stabilize and sustain a non-vanishing spin-polarization, or, magnetic moment, for a wide range of temperatures. Special focus was paid to compounds comprising localized spin moments, e.g., molecules and molecular aggregates containing transition metal or rare earth elements. The nuclear vibrations generate an effective field acting on the spin moments provided that the vibrations are anharmonic, which is typical for low symmetry structures, e.g., absence of inversion and/or reflection symmetries. The influence from this induced field increases in strength with temperature which, therefore, enables the vibrations to maintain a stable magnetic state also when the spin exchange is too small to provide this. The nature of the spin-exchange, whether it favors a ferromagnetic or anti-ferromagnetic state, becomes irrelevant whenever the temperature is sufficiently large for the vibrationally induced field to dominate the state. It was, finally, argued that the proposed mechanism for stabilizing the magnetic state gives a good qualitative agreement with recent experimental observations of, e.g., with temperature increasing coercivity accompanied with decreasing magnetization. Further experiments that would tie the unusual magnetic properties with, for instance, thermal properties, would be desired to facilitate further developments of the proposed theory.

\acknowledgements
Fruitful discussions with Anders Bergman, Ron Naaman, Lars Nordstr\"om, and Patrik Thunstr\"om are acknowledged. The authors thank Vetenskapsr\aa det for financial support.

\bibliography{CPISPref}

\end{document}